\documentstyle[12
pt,graphicx,caption2]{article}

\begin{document}

\title{On the critical examination of the "field-theoretical approach" to the neutron-antineutron oscillations in nuclei}
\author{V.I.Nazaruk\\
Institute for Nuclear Research of RAS, 60th October\\
Anniversary Prospect 7a, 117312 Moscow, Russia.}

\date{}
\maketitle
\bigskip

\begin{abstract}
We briefly outline the models of the $n\bar{n}$ transition in nuclei and analyse a well known calculation based on diagram technique published in 1992 and repeated recently by V. Kopeliovich and I. Potashnikova.
\end{abstract}

\vspace{5mm}

\vspace{5mm}

\vspace{1cm}

*E-mail: nazaruk@inr.ru

\newpage
\setcounter{equation}{0}

{\bf 1.} At present several models of $n\bar{n}$ transitions in medium are treated. The part of them gives radically different results. This is because the $n\bar{n}$ transition in nuclei is extremely sensitive to the details of the model and so we focus on the physics of the problem.

As far back as 1992, the model of $n\bar{n}$ transitions in light nuclei based on diagram technique for direct reactions (later on referred to as the model 1) has been published [1,2]. In 1996 this calculation was repeated for deuteron [3]. However, in [4] we abandoned this model for reasons given below. The model based on field-theoretical approach with finite time interval (model 2) has been proposed [4-6].

"Critical examination of the field-theoretical approach to the neutron-antineutron oscillations in nuclei" [7] is that the calculation of
[1-3] has been repeated. Also several "objections" against [4-6] are adduced. They are wrong or irrelevant (see section 2). For instance, in the abstract [7] we read: 

1) "Infrared divergences do not appear within the correct treatment of analytical properties of the amplitudes."

The analytical properties of the amplitudes have nothing to do with the infrared divergences. The infrared divergences do not appear since the model 1 used by authors [7] is, by construction, infrared-free (see section 3). The analytical properties of the amplitudes are related to
the calculation of the integral only (which is trivial).

In this paper we compare the model based on the field-theoretical approach (section 2) with the model based on diagram technique for direct reactions (section 3) and show that the latter model is unsuitable for the problem under study. Also it is shown that interpretation of a number of important points of the above-mentioned models given in [7] is wrong (sections 2-4). Section 5 contains the conclusion.

{\bf 2.} Model 2 describes both $n\bar{n}$ transition in the medium [4,5] (Fig. 1a) and nuclei [8] (Fig. 1b). It corresponds to the standard formulation of the problem: the $|in>$-state is the eigenfunctions of unperturbed Hamiltonian. In the case of the process shown in Fig. 1a, this is the neutron plane wave:
\begin{equation}
n_p(x)=\Omega ^{-1/2}\exp (-ipx),   
\end{equation}
$p=(\epsilon _n,{\bf p}_n)$, $\epsilon _n={\bf p}_n^2/2m+U_n$, where $U_n$ is the neutron potential. In the case of Fig. 1b, this is the wave function of bound state [8]. For the nucleus in the initial state we take the one-particle shell model. Since the neutron is described by the bound state wave function and not the plane wave, the virtual antineutron in Fig. 1b is described by Green function and not propagator. (The Feynman propagator is the partial case of Green function [8,9].)

We consider the process shown in Fig. 1a. The interaction Hamiltonian is
\begin{eqnarray}
{\cal H}_I={\cal H}_{n\bar{n}}+{\cal H},\nonumber\\
{\cal H}_{n\bar{n}}=\epsilon \bar{\Psi }_{\bar{n}}\Psi _n+H.c.
\end{eqnarray}
Here ${\cal H}_{n\bar{n}}$ and ${\cal H}$ are the Hamiltonians of the $n\bar{n}$ transition [10] and the $\bar{n}$-medium interaction, respectively; $\epsilon $ is a small parameter with $\epsilon =1/\tau $, where $\tau $ is the free-space $n\bar{n}$ 
oscillation time; $\Psi _n$ and $\Psi _{\bar{n}}$ are the operators of the neutron and antineutron fields; $m_n=m_{\bar{n}}=m$. 

\begin{figure}[h]
  {\includegraphics[height=.25\textheight]{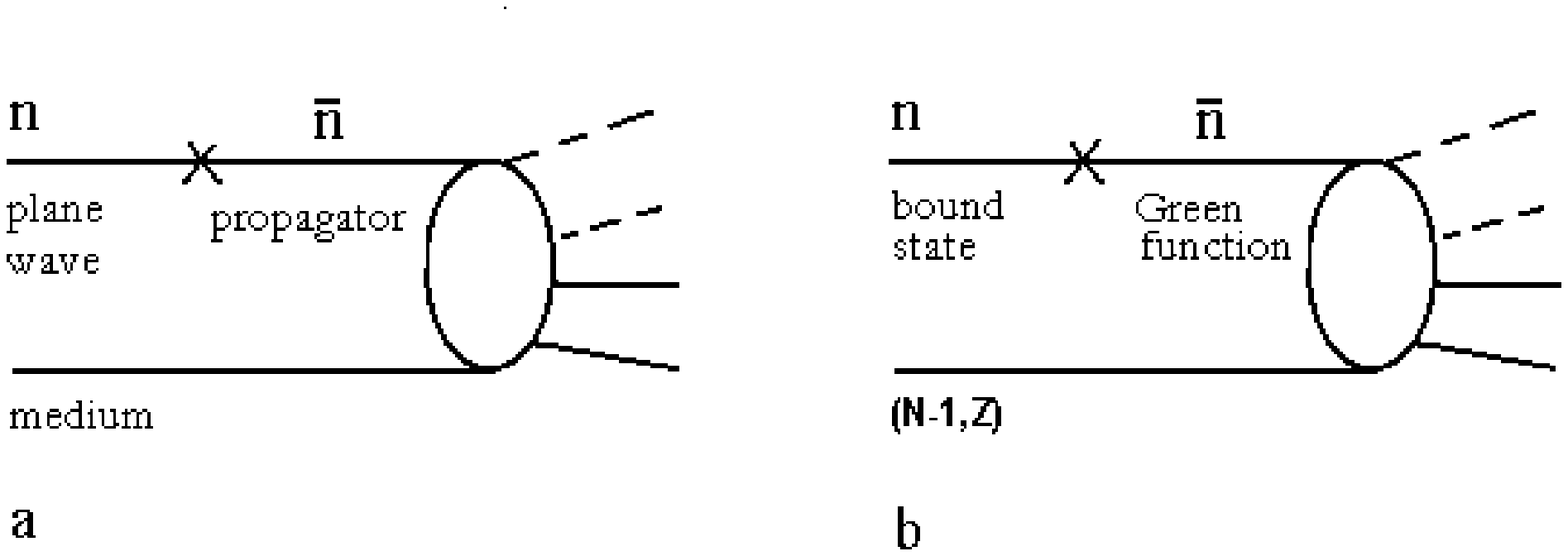}}
  \caption{$n\bar{n}$ transition in the medium ({\bf a}) and nuclei ({\bf b}) followed by annihilation.}
\end{figure}

In the model 2 the antineutron propagator can be bare or dressed. The corresponding results differ radically. If the antineutron propagator is bare  (the antineutron self-energy $\Sigma =0$), the $S$-matrix amplitudes corresponding to Figs. 1a and 1b contain the infrared singularity. It is conditioned by zero momentum transfer in the $n\bar{n}$ transition vertex. To avoid the infrared singularities the problem is formulated on the finite time interval $(t,0)$ [6]. Then the matrix element of evolution operator $U(t,0)$ is calculated (see Eq. (13) of Ref. [4]). The lower limit on the free-space $n\bar{n}$ oscillation time is found to be [4,5]
\begin{equation}
\tau ^b=10^{16}\; {\rm yr}.
\end{equation}
This value is interpreted as the estimation from above.

The fact that the evolution operator is used is not new [9,11]. Recall that the most part of physical problems is formulated on the finite time interval. The model described above is standard and so it created no questions up to now. However, in [7] we read:

2) "The author of [4-6] tries to reconstruct the space-time picture of the process..."

In reality we do not more than calculate the matrix element of evolution operator.

3) "...new rules seem to be proposed in [4,5] instead of well known Feynman rules."

We calculate the matrix element of evolution operator only. In regard Feynman rules, the problem formulation (process model) in the quantum field theory and nuclear physics differ principally. In particular, there is density-dependence (coordinate-dependence) in nuclear physics. This is absolutely different problems and so the Feynman rules developed in quantum electrodynamics (QED) can be only element of the corresponding model of nuclear reaction or decay. This model (more precisely, the model of $n\bar{n}$ transitions in light nuclei based on diagram technique for direct reactions) was published in [1,2] and analyzed in the next section. So the words "instead of well known Feynman rules" mislead the reader at least. 

4) "These new rules should allow to reproduce all the well known results of nuclear theory". 

The consideration of the matrix element of evolution operator is interpreted [7] as "new rules". As for the second part of the remark, indeed, all the well known results of nuclear theory have been obtained by means of $S$-operator ($U$-operator).

The "remarks" 1) - 4) are due to the misunderstanding of the authors of [7]. Also we recall that the above-given problem formulation created no questions up to now. If the antineutron self-energy equal to zero, the process amplitude is singular. In this case the {\em calculation} of matrix element is really non-standard [4,5]. However, the pp. 1)-4) are irrelevant to the calculation of matrix element.

For the model 2 with dressed propagator the calculation is standard [5,13]. The lower limit on the free-space $n\bar{n}$ oscillation time is
\begin{equation}
\tau ^d=1.2\cdot 10^{9}\; {\rm s}.
\end{equation}
The models with bare and dressed propagators were analyzed and compared in [5].

{\bf 3.} Model 1 is shown in Fig. 2. Denote: $A=(N,Z)$, $B=(N-1,Z)$ are the initial and intermediate nuclei, $M$ is the amplitude of virtual decay $A\rightarrow n+(A-1)$, $M_a^{(n)}$ is the amplitude of $\bar{n}B$ annihilation in $(n)$ mesons, $E_n$ is the pole neutron binding energy; $m$, $m_A$, $m_B$ are the masses of the nucleon and nuclei $A$ and $B$, respectively. The process amplitude is [2]
\begin{equation}
M^{(n)}=-\frac{im^2m_B}{2\pi ^4}\epsilon 
\int d{\bf q}dE \frac{M(q)M_a^{(n)}(m_A)}{({\bf q}^2-2mE-i0)^2[{\bf q}^2+2m_B
(E+E_n)-i0)]}.
\end{equation}

\begin{figure}[h]
  {\includegraphics[height=.25\textheight]{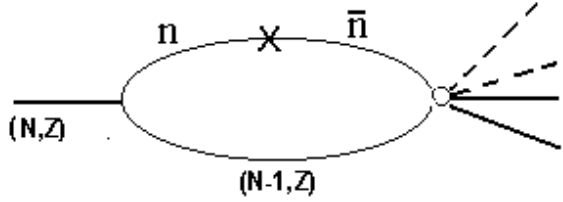}}
  \caption{Model 1 for the $n\bar{n}$ transition in the nuclei followed by annihilation.}
\end{figure}

For deuteron the process probability $W_1(t)$ is given by Eqs. (6), (12) and (18) of Ref. [2]:
\begin{eqnarray}
W_1(t)=\Gamma _{d\rightarrow mesons}t,\nonumber\\
\Gamma _{d\rightarrow mesons}=\frac{\epsilon ^2}{6E_n^2}\Gamma _{\bar{n}p},
\end{eqnarray}
where $\Gamma _{\bar{n}p}$ is the $\bar{n}p$ annihilation width. This is final result. At the same time the authors [7] write "There is no final formula for $\Gamma _{d\rightarrow mesons}$ in [2]."

We list the main drawbacks of the above-given model which are essential for the problem under study:

a) The model does not reproduce the $n\bar{n}$ transitions in the medium and vacuum. If $E_n\rightarrow 0$, $W_1$ diverges (see also Eqs. (15) and (17) of Ref. [3]). 

b) Contrary to the model 2, the amplitude (3) cannot be obtained from the Hamiltonian because in the interaction Hamiltonian {\em there is no term} which induces the virtual decay $(N,Z)\rightarrow n+(N-1,Z)$.

c) The model does not contain the infrared singularity for any process including the $n\bar{n}$ transition, whereas it exists for the processes in the medium and vacuum (see [4-6]). This brings up the question: Why? The answer is that for the propagator the infrared singularity cannot be in principle since the particle is virtual: $p_0^2\neq m^2+{\bf p}^2$. Due to this the model is infrared-free. On the other, hand the neutron is in the bound state and should be described by wave function and not the propagator.

d) Since the model is formulated in the momentum representation, it does not describe the coordinate-dependence, in particular the loss of particle intensity due to absorption. Also there is a no the dependence on nuclear density! The model is crude and has very restricted range of applicability.  

We consider the points b) and c). The $n\bar{n}$ transition takes place in the propagator. As the result the model is infrared-free. For the processes with zero momentum transfer this fact is crucial since it changes the functional structure of the amplitude.

On the other hand, the neutron propagator arises owing to the vertex of virtual decay $A\rightarrow n+(A-1)$. However, as pointed out above, in the interaction Hamiltonian {\em there is no term} which induces the virtual decay $A\rightarrow n+(A-1)$. This vertex is the artificial element of the model. It was introduced in order for the neutron (pole particle) state to be separated.

We assert that for the problem under study this scheme is incorrect. The neutron state is described wrongly. The diagram technique for direct reactions has been developed and adapted to the direct type reactions. The term "diagram technique for direct reactions" emphasizes this circumstance. The processes with non-zero momentum transfer are considerably less sensitive to the description of pole particle state. The approach is very handy, useful and simple since it is formulated in the momentum representation. It was applied by us for the calculation of knock-out reactions and $\bar{p}$-nuclear annihilation [12]. The price of simplicity is that its applicability range is {\em restricted}. At the same time, as is seen from p. c), the process under study is extremely sensitive to the description of neutron state. The same is true for the value of antineutron self energy [13]. Besides, the problem is unstable [13].

Finally, since the operator ${\cal H}_{n\bar{n}}$ acts on the neutron, in the model 1 the vertex of virtual decay $A\rightarrow n+(A-1)$ is introduced because one should separate out the neutron state. This scheme is artificial because in the interaction Hamiltonian {\em there is no term} which induces the virtual decay $A\rightarrow n+(A-1)$. The neutron of the nucleus is in the bound state and so it cannot be described by the propagator. More precisely the model containing the vertex of virtual decay is a crude one. It is inapplicable for the problem under study. Alternative method is given by the model 2 which does not contain the above-mentioned vertex. The shell model used in the $|in>$-state of the model 2 has no need of a commentary. 

For above-given reasons we abandoned this model [4]. The model based on the potential description of $\bar{n}$-medium interaction (potential model) has been considered in [5,14] in detail.

{\bf 4.} The rest of the assertions of [7] should be commented.

5) The words "we give some general arguments..." are wrong since these arguments result from the model 1.

Figure 1 adduced in [7] determines uniquely the process model (model 1). Equation (8) and subsequent consideration relate to the model 1.

6) The consideration of "energy scale" [7] is nonsense. What is the "energy scale" for nuclear $\beta $-decay, for example?

7) "The reason of suppression is the localization of the neutron inside the nucleus."

To understand that the localization of the neutron inside the nucleus is unrelated to the suppression, it is suffice to compare the nuclear $\beta $-decay and decay of free neutron. The sole possible reason discussed in the literature [14,15] is the antineutron potential. In the
models 1 and 2 the potential is not introduced and above-mentioned mechanism of suppression is inoperative. We also recall that the $n\bar{n}$ transition in the nuclei followed by annihilation is the dynamical, two-step process with the characteristic time $\tau _{ch}\sim 10^{-24}$ s [4,5,13].  

8) "If the infrared divergence takes place for the process of $n\bar{n}$ transitions in nucleus, it would take place also for the nucleus form-factor at zero momentum transfer. But it is well known not to be the case, as we illustrated in this section"

So the authors illustrated that the model 1 is infrared-free and the large part of [7] is devoted to this "problem". However, this trivial fact is well known [9,11]: the infrared divergence cannot take place for virtual particle. 

The errors indicated in pp. 1) - 8) and the fact that the main calculation is the special case of the model proposed by us [1,2] in 1992 
are incompatible with the paper title "Critical examination..." and the beginning of the acknowledgements "Present investigation, performed partly with {\em pedagogical} purposes".

{\bf 5.} In conclusion, the process under study is extremely sensitive to the description of neutron state (model 1 or 2). The same is true for the value of antineutron self-energy $\Sigma $. If $\Sigma $ changes in the limits $10\; {\rm MeV}>\Sigma >0\; {\rm MeV}$, the lower limit on the free-space $n\bar{n}$ oscillation time $\tau $ is in the range [8,13]

\begin{equation}
10^{16}\; {\rm yr}>\tau >1.2\cdot 10^{9}\; {\rm s}.
\end{equation}
For the free-space $ab$ oscillations the picture is exactly the same: the $ab$ transition probability is extremely sensitive to the difference of masses $m_a-m_b$. The fact that process amplitude is in the peculiar point (see Fig. 1) is the basic reason why the range (7) is very wide. The values $\tau ^d=1.2\cdot 10^{9}\; {\rm s}$ and $\tau ^b= 10^{16}$ yr are interpreted as the estimations from below (conservative limit) and from above, respectively. The estimation from below $\tau ^d=1.2\cdot 10^{9}\; {\rm s}$ exceeds the restriction given by the Grenoble reactor experiment [16] by a factor of 14 and the lower limit given by potential model [17] by a factor of 5. At the same time the range of uncertainty of $\tau $ is too wide. Further theoretical and experimental investigations are desirable.

\end{document}